\documentclass[12pt]{article}
\usepackage{amssymb}

\usepackage{graphicx}
\usepackage{amsmath}

\begin{document}

\title{FIVE-DIMENSIONAL RELATIVITY AND TWO TIMES}
\date{}
\author{}
\maketitle

\begin{center}
\bigskip Paul S. Wesson$^{1,\,2}$
\end{center}

\begin{enumerate}
\item \bigskip Dept. Physics, University of Waterloo, Waterloo, Ontario N2L
3G1, Canada \linebreak Keywords: Time, 5D Relativity, Waves, Equivalence
Principle

Pacs: 05.70.Ln, 04.50.+h, 0.4.20.Cv

\item Correspondence: mail = (1) above; fax=(519) 746-8115;

email=wesson@astro.uwaterloo.ca. \pagebreak
\end{enumerate}

\begin{center}
{\LARGE FIVE-DIMENSIONAL RELATIVITY AND TWO TIMES}
\end{center}

\underline{{\LARGE Abstract}}

It is possible that null paths in 5D appear as the timelike paths of massive
particles in 4D, where there is an oscillation in the fifth dimension around
the hypersurface we call spacetime. A particle in 5D may be regarded as
multiply imaged in 4D, and the 4D weak equivalence principle may be regarded
as a symmetry of the 5D metric.

\section{\protect\underline{Introduction}}

In quantum theory, it has recently been shown that the statistical
interactions of particles can lead to thermodynamic arrows of time for
different parts of the universe which are different or even opposed [1]. In
relativity theory, it is well known how to incorporate the phenomenological
laws of standard thermodynamics into general relativity, but it has not been
clear how to treat particle dynamics and the nature of time in
higher-dimensional manifolds which may unify quantum theory and gravity.
However, a couple of exact solutions of the field equations of 5D relativity
have recently been found which have good physical properties but involve
manifolds with signature $\left[ +\left( ---\right) +\right] $ that describe
two ``time'' dimensions [2]. Such two-time metrics are currently the subject
of investigation in relation to even higher-dimensional extensions of
general relativity, notably string theory [3]. While field theory in $N$
dimensions is an alluring subject, concrete calculations which might be
compared to observations are unfortunately compromised somewhat by
uncertainty about $N\;(4<N\leq 26$ is a commonly-held view). The case $N=5$
continues to be a major focus, because it is the basic extension of general
relativity and the low-energy limit of more extended theories. Thus $N=5$
induced-matter theory and membrane theory are widely regarded as the best
options for resolving the cosmological-constant and hierarchy problems for
the energy density of the vacuum and masses of particles, because they both
drop the hobbling cylinder condition (no dependence on the extra coordinate)
typical of old Kaluza-Klein theory. The versatility of 5D relativity has
recently been illustrated again by the demonstration that timelike paths in
4D can be interpreted as \underline{null} paths in 5D [4, 5]. This implies
that what we regard as massive particles moving through spacetime, with
finite separations in ordinary 3D space and 4D proper time, are photon-like
objects in 5D with no ``separation'' in supertime. Such an interpretation
should not be viewed as merely a neat concept, because if we use the 4D
proper time of conventional relativity to parametize the motion of a
particle in 5D, the geodesic equations for the latter show that in general
there is a fifth force which acts parallel to the 4-velocity in both
induced-matter and membrane theory and is in principle open to observation
[6, 7]. The implications of this are major, irrespective of whether we live
in a world where matter is the manifestation of an unconstrained fifth
dimension [2, 4, 6, 8] or in one where we are constrained to a hypersurface
in a shadowy five-dimensional ``bulk'' [3, 5, 7, 9]. The two approaches
appear to be equivalent [10], at least mathematically.

In what follows we wish to extend what has been noted above and derive
several new results in two-time 5D relativity that are remarkable.

\section{\protect\underline{Properties of Null 5D Two-Time Metrics}}

By Campbell's theorem, it is always possible to embed a 4D Riemannian line
element with $ds^{2}=g_{\alpha \beta }dx^{\alpha }dx^{\beta }$ in a 5D one
with $dS^{2}=g_{AB}dx^{A}dx^{B}$ (here $g_{\alpha \beta }$ and $g_{AB}$ are
the 4D and 5D metric tensors with $\alpha =0,123$ and $A=0,123,4$). However,
$g_{AB}=g_{AB}\left( x^{\alpha },\,l\right) $ where $x^{4}=l$ is the extra
coordinate, so 5D quantities calculated from $g_{AB}$ will in general be $%
Q=Q\left( x^{\alpha },l\right) $. The latter behave covariantly under the
group of 5D coordinate transformations $x^{A}\rightarrow \overline{x}%
^{A}\left( x^{B}\right) $, but \underline{not} under the conventional group
of 4D transformations $x^{\alpha }\rightarrow \overline{x}^{\alpha }\left(
x^{\beta }\right) $. Thus a choice of 5D coordinates (or gauge) is necessary
in order to specify 4D physics.

In the quasi-Minkowski gauge, a particle moving along a null path in a
two-time 5D metric has
\begin{equation}
0=dS^{2}=dt^{2}-\left( dx^{2}+dy^{2}+dz^{2}\right) +dl^{2}\;\ \ .
\end{equation}%
The 5-velocities $U^{A}\equiv dx^{A}\diagup d\lambda $ where $\lambda $ is
an affine parameter obey $U^{A}U_{A}=0$ (we absorb the speed of light $c$
here, and the gravitational constant $G$ and Planck's constant $h$
elsewhere, by a choice of units that renders them all unity). With $\lambda
=s$ for the proper 4D time, the velocity in ordinary space $\left( v\right) $
is related to the velocity along the axis of ordinary time $\left( u\right) $
and the velocity along the fifth dimension $\left( w\right) $ by $%
v^{2}=u^{2}+w^{2}$. This implies super-luminal speeds; but the particle
which follows the path specified by (1) should not be identified with the
tachyon of special relativity, because in both induced-matter [2, 4] and
brane theory [5, 7] $l$ is related to the rest mass of a test particle.

In the non-electromagnetic gauge, the line element can be written%
\begin{equation}
dS^{2}=g_{\alpha \beta }\left( x^{\gamma },\,l\right) dx^{\alpha }dx^{\beta
}+\epsilon \;\Phi ^{2}\left( x^{\gamma },\;l\right) \;\;\;.
\end{equation}%
This uses only 4 of the 5 available degrees of coordinate freedom to remove
the electromagnetic potentials $\left( g_{4\alpha }\right) $, leaving the
scalar or Higgs potential and the signature general $\left( g_{44}=\epsilon
\;\Phi ^{2}\right) $. The gravitational potentials $\left( g_{\alpha \beta
}\right) $ are also general. The components of the 5D Ricci tensor for the
metric (2) deserve to be generally known, and we therefore tabulate them in
what we believe to be their most convenient form:
\begin{eqnarray}
^{5}R_{\alpha \beta } &=&\,^{4}R_{\alpha \beta }-\frac{\Phi _{,\alpha ;\beta
}}{\Phi }+\frac{\epsilon }{2\Phi ^{2}}\left( \frac{\Phi _{,4}g_{\alpha \beta
,4}}{\Phi }-g_{\alpha \beta ,44}\right.   \notag \\
&&\left. +g^{\lambda \mu }g_{\alpha \lambda ,4}g_{\beta \mu ,4}-\frac{g^{\mu
\nu }g_{\mu \nu ,4}g_{\alpha \beta ,4}}{2}\right)
\end{eqnarray}%
\begin{eqnarray}
R_{44} &=&-\epsilon \Phi \square \Phi -\frac{g_{\;\;\;,4}^{\lambda \beta
}g_{\lambda \beta ,4}}{2}-\frac{g^{\lambda \beta }g_{\lambda \beta ,44}}{2}+%
\frac{\Phi _{,4}g^{\lambda \beta }g_{\lambda \beta ,4}}{2\Phi }  \notag \\
&&\;\;\;\;\;\;\;\;\;\;\;\;\;\;\;\;\;\;\;\;\;\;\;-\frac{g^{\mu \beta
}g^{\lambda \sigma }g_{\lambda \beta ,4}g_{\mu \sigma ,4}}{4}
\end{eqnarray}%
\begin{eqnarray}
R_{4\alpha } &=&\Gamma \left( \frac{g^{\beta \lambda }g_{\lambda \alpha
,4}-\delta _{\alpha }^{\beta }g^{\mu \nu }g_{\mu \nu ,4}}{2\Gamma }\right)
_{,\beta }  \notag \\
&&+\frac{g^{\mu \beta }g_{\mu \beta ,\lambda }g^{\lambda \sigma }g_{\sigma
\alpha ,4}}{4}-\frac{g^{\lambda \beta }g_{\beta \mu ,\alpha }g^{\mu \sigma
}g_{\sigma \lambda ,4}}{4}\;\;\;.
\end{eqnarray}%
Here a comma denotes the ordinary partial derivative, a semicolon denotes
the ordinary 4D covariant derivative, $\square \Phi \equiv g^{\mu \nu }\Phi
_{,\mu ;\nu }$ and $\Gamma \equiv \left| \epsilon \Phi ^{2}\right| ^{1/2}$.
The field equations of 5D relativity are commonly taken to be $R_{AB}=0$.
Then (4) is a wave equation for the scalar field, (5) can be couched as a
set of 4 conservation equations, and (3) can be put into the form of the 10
Einstein equations [2, 4, 10]. The latter read $G_{\alpha \beta }=8\pi
\;T_{\alpha \beta },$ where the Einstein tensor $G_{\alpha \beta }\equiv
\;^{4}R_{\alpha \beta }-\;^{4}Rg_{\alpha \beta }\diagup 2$ is constructed
from the remaining 5D quantities. Using $R_{AB}=0$ in (3) - (5), the 4D
scalar curvature is%
\begin{equation}
^{4}R=\frac{\epsilon }{4\Phi ^{2}}\left[ g_{\;\;\;,4}^{\mu \nu }\;g_{\mu \nu
,4}+\left( g^{\mu \nu }\;g_{\mu \nu ,4}\right) ^{2}\right] \;.
\end{equation}%
This relation has been used implicitly in the literature, but explicitly as
here it is extremely instructive: (a) what we call the curvature of 4D
spacetime can be regarded as the result of embedding it in an $x^{4}$%
-dependent 5D manifold; (b) the sign of the 4D curvature depends on the
signature of the 5D metric, which here is $+(---)\pm $ and admits the
two-time option; (c) the magnitude of the 4D curvature depends strongly on
the scalar field or the size of the extra dimension $\left( g_{44}=\epsilon
\Phi ^{2}\right) $, so while it may be justifiable to neglect this in
astrophysics (where the 4D curvature is small) it can be crucial in
cosmology and particle physics. The 4D energy-momentum tensor that follows
from (3)-(5) is
\begin{eqnarray}
8\pi T_{\alpha \beta } &=&\frac{\Phi _{,\alpha ;\beta }}{\Phi }-\frac{%
\epsilon }{2\Phi ^{2}}\left\{ \frac{\Phi _{,4}g_{\alpha \beta ,4}}{\Phi }%
-g_{\alpha \beta ,44}+g^{\lambda \mu }g_{\alpha \lambda ,4}g_{\beta \mu
,4}\right.   \notag \\
&&\left. -\frac{g^{\mu \nu }g_{\mu \nu ,4}g_{\alpha \beta ,4}}{2}+\frac{%
g_{\alpha \beta }}{4}\left[ g_{\;\;,4}^{\mu \nu }g_{\mu \nu ,4}+\left(
g^{\mu \nu }g_{\mu \nu ,4}\right) ^{2}\right] \right\} \;\;\;.
\end{eqnarray}%
The existence of this relation has been inferred in the literature as a
corollary of Campbell's theorem, but explicitly as here it is also extremely
instructive: \ (a) what we call matter in a curved 4D spacetime can be
regarded as the result of the embedding in an $x^{4}$-dependent (possibly
flat) 5D manifold; (b) the nature of the 4D matter depends on the signature
of the 5D metric; (c) the 4D source depends on the extrinsic curvature of
the embedded 4D spacetime and the scalar field associated with the extra
dimension, which while they are in general mixed correspond loosely to
ordinary matter and the stress-energy of the vacuum.

In the pure-canomical gauge of induced-matter theory, the first part of (2)
is factorized via $g_{\alpha \beta }\left( x^{\gamma },\;l\right) =\left(
l^{2}\diagup L^{2}\right) \overline{g}_{\alpha \beta }\left( x^{\gamma
}\right) $ where $L$ is a constant length, and the last part of (2) is fixed
via $g_{44}=-1$. There is a considerable literature on this gauge [2, 4],
which is related to the warp gauge of membrane theory [5, 7: the latter uses
a factor that is exponential in $l$]. Then (7) causes Einstein's equations
to read $G_{\alpha \beta }=-3\epsilon \overline{g}_{\alpha \beta }\diagup
L^{2}$, which defines a cosmological constant $\Lambda =-3\epsilon \diagup
L^{2}$. [Relation (6) also gives the standard relation for an embedded
vacuum spacetime.] The cosmological constant can, of course, be regarded as
defining the energy density and pressure of the vacuum in Einstein's theory
via $\rho =-p=\Lambda \diagup 8\pi $. This is a special case of (7), and we
now turn our attention to the general relations (3)-(7) with an extra
timelike dimension.

A simple, wave-like solution of $R_{AB}=0$ is given by
\begin{equation}
dS^{2}=\frac{l^{2}}{L^{2}}\left[ dt^{2}-e^{i\left( \omega t+k_{x}x\right)
}dx^{2}-e^{i\left( \omega t+k_{y}y\right) }dy^{2}-e^{i\left( \omega
t+k_{z}z\right) }dz^{2}\right] +dl^{2}.
\end{equation}%
Here $k_{xyz}$ are wave numbers and the frequency is constrained by the
solution to be $\omega =\pm 2\,\diagup \,L$. We have studied (8) both
algebraically and computationally using the program GRTensor (which may also
be used to verify it). Solution (8) clearly has two ``times''. It also has
complex metric coefficients for the ordinary 3D space, but closer inspection
shows that the structure of the field equations leads to physical quantities
that are real. The 3D wave is not of the sort found in general relativity,
but owes its existence to the choice of coordinates. A trivial change in the
latter suppresses the appearance of the wave in 3D space, in analogy to how
a transverse wave is noticed or not by an observer, depending on whether he
is fixed in the laboratory frame or moving with the wave. A further change
of coordinates can be shown to make (8) look like the 5D analog of the de
Sitter solution. This leads us to conjecture that the wave is supported by
the pressure and energy density of a vacuum with the equation of state found
in general relativity, namely $p+\rho =0$. This is confirmed to be the case,
with $\Lambda <0.$ It may also be confirmed that (8) is not only Ricci-flat $%
\left( R_{AB}=0\right) $ but also Riemann-flat $\left( R_{ABCD}=0\right) .$
It is a wave travelling in a curved 4D spacetime that is embedded in a
\underline{flat} 5D manifold which has no energy.

The logical condition on the path, for a particle moving in a 5D manifold
that has no energy, is that it be null. To make contact with other work we
choose a two-time metric of canonical form, so
\begin{equation}
0=dS^{2}=\frac{l^{2}}{L^{2}}ds^{2}+dl^{2}\;\;\;.
\end{equation}%
Here we take $ds^{2}=\overline{g}_{\alpha \beta }\left( x^{\alpha },l\right)
dx^{\alpha }dx^{\beta },$ using all of the 5 available coordinate degrees of
freedom to suppress the potentials of electromagnetic and scalar type, but
leaving the metric otherwise general. The solution of (9) is $l=l_{0}\exp %
\left[ \pm i\left( s-s_{0}\right) \diagup L\right] $, where $l_{0}$ and $%
s_{0}$ are constants of which the latter may be absorbed. Then $%
l=l_{0}\;e^{\pm is\diagup L}$ describes an $l$-orbit that oscillates about
spacetime with amplitude $l_{0}$ and wavelength $L$. The motion is actually
simple harmonic, since $d^{2}l\diagup ds^{2}=-l\diagup L^{2}$. Also, $%
dl\diagup ds=\pm il\diagup L$, so the physical identifications of the mass
of a particle with the momentum in the extra dimension as in brane theory
[5, 7] or with the extra coordinate as in induced-matter theory [2, 4] are
equivalent, modulo a constant. In both cases, \underline{the $l$-orbit
intersects the $s$-plane an infinite number of times}. There is only one
period in the metric (9), defined by $L$, but of course a Fourier sum of
simple harmonics can be used to construct more complicated orbits in the $%
l\diagup s$ plane. [Alternatively, extra length scales can be introduced to
(9) via $L^{2}\rightarrow \left( L_{1}^{2}L_{2}^{2}L_{3}^{2}\ldots
L_{n}^{2}\right) \diagup \left( L_{2}^{2}L_{3}^{2}\ldots
L_{n}^{2}+L_{1}^{2}L_{3}^{2}\ldots L_{n}^{2}+\ldots \right) $.] If we
identify the orbit in the $l\diagup s$ plane with that of a particle, we
have a realization of the old idea (often attributed to Wheeler and/or
Feynmann) that instead of there being 10$^{80}$ particles in the visible
universe there is in fact only one which appears 10$^{80}$ times.

The above description is classical, but there is no impediment to its
extension to the quantum domain [2]. By (9), the traditional sum over $s$%
-paths in 4D can if so desired be replaced by a sum over $l$-paths in 5D.
However, metrics like (9) still require that the 5D path $S$ be minimized
around zero. The condition for this is given by the 5D geodesic equation,
which for (9) can conveniently be presented in terms of the equation of
motion in spacetime and the equation of motion in the extra dimension:
\begin{eqnarray}
\frac{du^{\mu }}{ds}+\Gamma _{\beta \gamma }^{\mu }u^{\beta }u^{\gamma }
&=&f^{\mu } \\
f^{\mu } &\equiv &\left( -g^{\mu \alpha }+\frac{u^{\mu }u^{\alpha }}{2}%
\right) u^{\beta }\frac{dl}{ds}\frac{\partial g_{\alpha \beta }}{\partial l}
\\
\frac{d^{2}l}{ds^{2}}-\frac{2}{l}\left( \frac{dl}{ds}\right) ^{2}-\frac{l}{%
L^{2}} &=&\frac{1}{2}\left\{ \frac{l^{2}}{L^{2}}+\left( \frac{dl}{ds}\right)
^{2}\right\} u^{\alpha }u^{\beta }\frac{\partial g_{\alpha \beta }}{\partial
l}\;\;\;.
\end{eqnarray}%
Here $u^{\mu }\equiv dx^{\mu }\diagup ds$, $\Gamma _{\beta \gamma }^{\mu }$
is the usual 4D Christoffel symbol, and $f^{\mu }$ is the fifth force (per
unit inertial mass) which has been discussed for induced-matter theory [6]
and membrane theory [7]. By (10), the motion in spacetime is only geodesic
in the usual 4D sense if $f^{\mu }=0$. This force may be split into a part $%
\left( -g^{\mu \alpha }+u^{\mu }u^{\alpha }\right) u^{\beta }\left(
dl\diagup ds\right) \left( \partial g_{\alpha \beta }\diagup \partial
l\right) $ which by construction is normal to the 4-velocity $u^{\mu },$ and
a part -$\left( u^{\mu }\diagup 2\right) \left( u^{\alpha }u^{\beta
}\partial g_{\alpha \beta }\diagup \partial l\right) \left( dl\diagup
ds\right) $ which is parallel to it. The latter has no analog in 4D general
relativity or any field theory like it, including electromagnetism. The
question of why the fifth force has hitherto not been observed is therefore
tantamount to the question of why the scalar quantity $Q\equiv u^{\alpha
}u^{\beta }\left( \partial g_{\alpha \beta }\diagup \partial l\right) $ is
small. [Clearly $u^{\mu }\neq 0$ in general, $dl\diagup ds=0$ would
effectively reduce the metric from 5D to 4D, and the null wave $%
l=l_{0}e^{\pm is\diagup L}$ satisfies (12) with no constraint on $Q$.] Our
answer to this is as follows: $Q=0$ if $\partial g_{\alpha \beta }\diagup
\partial l=0$, meaning that there is no intrusion of the fifth dimension
into spacetime, irrespective of the physical identification of $x^{4}=l$.
This says that the 4D weak equivalence principle is a symmetry of the 5D
metric.

\section{\protect\underline{Conclusion}}

There are non-unique times in 4D statistical mechanics [1], 5D relativity
[2] and $N\left( >5\right) D$ string theory [3]. In 5D, particles which are
dynamically massless can appear to be massive in spacetime [4, 5], and are
acted upon in general by a fifth force [6, 7] which can manifest itself in
both induced-matter and membrane theory [8, 9], though these approaches are
mathematically equivalent [10]. In the present work, we have outlined the
main consequences of 5D relativity with signature $\left[ +\left( ---\right)
+\right] $. Though technically referred to as ``two-time'' metrics, there is
no problem with closed timelike paths because the second timelike coordinate
is related to the (inertial) rest mass of a particle in both induced-matter
and membrane theory [2, 7]. To derive physical results requires the
specification of a gauge, as in (1) or (2). However, we have given the
components of the 5D Ricci tensor $R_{AB}$ for a general non-electromagnetic
gauge in (3)-(5). These for field equations $R_{AB}=0$ lead to expressions
for the embedded 4D Einstein space, namely (6) for the curvature scalar and
(7) for the energy-momentum tensor. The exact solution (8) illustrates the
new physics deriveable from two-time metrics: it describes a wave moving
through a de Sitter vacuum. The general two-time metrics (9) describe waves
in the fifth dimension that oscillate around the hypersurface we call
spacetime. This can be used as a model for multiply-imaged particles. The
dynamics of the latter are governed by relations (10)-(12), which show that
the fifth force typical of induced-matter and membrane theory is absent if
the weak equivalence principle is invoked as a symmetry of the metric.

\bigskip

\underline{{\LARGE Acknowledgements}}

This work grew out of earlier collaborations with H. Liu, B. Mashhoon and
S.S. Seahra. It was supported by N.S.E.R.C.

\bigskip

\underline{{\LARGE References}}

\begin{enumerate}
\item L.S. Schulman, Phys. Rev. D \underline{7}, 2868 (1973). L.S. Schulman,
Time's Arrow and Quantum Measurement (Cambridge Un. Press, Cambridge, 1997).
L.S. Schulman, Phys. Rev. Lett. \underline{83}, 5419 (1999). L.S. Schulman,
Phys. Rev. Lett. \underline{85}, 897 (2000).

\item P.S. Wesson, Space-Time-Matter (World Scientific, Singapore, 1999). A.
Billiard, P.S. Wesson, Phys. Rev. D \underline{53}, 731 (1996).A. Billiard,
P.S. Wesson, Gen. Rel. Grav. \underline{28}, 129 (1996). P.S. Wesson,
Observatory \underline{121}, 82 (2001). P.S. Wesson, J. Math. Phys. in press
(2002). (gr-gc/0105059, 2001.)

\item I. Bars, C. Kounnas, Phys. Rev. D \underline{56}, 3664 (1997). I.
Bars, C. Deliduman, O. Andreev, Phys. Rev. D \underline{58}, 066004 (1998).
I. Bars, C. Deliduman, D. Minic, Phys. Rev. D \underline{59}, 125004 (1999).
J. Kocinski, M. Wierzbicki, gr-gc/0110075 (2001).

\item S.S. Seahra, P.S. Wesson, Gen. Rel. Grav. \underline{33}, 1731 (2001).

\item D. Youm, hep-th/0110013 (2001).

\item P.S. Wesson, B. Mashhoon, H. Liu, W.N. Sajko, Phys. Lett. B \underline{%
456}, 34 (1999).

\item D. Youm, Phys. Rev. D \underline{62}, 084002 (2000).

\item P.S. Wesson, Phys. Lett. B \underline{276}, 299 (1992).

\item L. Randall, R. Sundrum, Mod. Phys. Lett. A \underline{13}, 2807 (1998).

\item J. Ponce de Leon, Mod. Phys. Lett. A \underline{16}, 2291 (2001).
\end{enumerate}

\end{document}